\documentstyle[12pt,epsfig]{article}
\begin{document}

\noindent Stockholm\\
USITP 06-01\\
January 2006\\

\vspace{1cm}

\begin{center}

{\Large FLAT INFORMATION GEOMETRIES}

\vspace{8mm}

{\Large IN BLACK HOLE THERMODYNAMICS}

\vspace{1cm}

{\large Jan E. \AA man}\footnote{Email address: ja@physto.se.}

\

{\large Ingemar Bengtsson}\footnote{Email address: ingemar@physto.se. 
Supported by VR.}

\

{\large Narit Pidokrajt}\footnote{Email address: narit@physto.se.}

\

{\sl Stockholm University, AlbaNova\\
Fysikum\\
S-106 91 Stockholm, Sweden}

\vspace{8mm}

{\bf Abstract}

\end{center}

\vspace{5mm}

\noindent The Hessian of either the entropy or the energy function 
can be regarded as a metric on a Gibbs surface. For two parameter 
families of asymptotically flat black holes in arbitrary dimension 
one or the other of these metrics are flat, and the state space is 
a flat wedge. The mathematical reason for this is traced back to 
the scale invariance of the Einstein-Maxwell equations. The 
picture of state space that we obtain makes some properties such as 
the occurence of divergent specific heats transparent. 

\newpage

{\bf 1. Introduction}

\

\noindent It has been known for a long time that black holes can be 
described as thermodynamic systems. The connection to ordinary 
thermodynamics was clinched by Hawking's calculation, showing 
that the surface gravity of the event horizon equals $2\pi$ times 
the ordinary temperature of the radiation emitted by the black hole 
to infinity \cite{Hawking}. The original derivation of the laws of black hole 
thermodynamics has nothing in common with statistical mechanics, but 
there is a general belief that a connection nevertheless exists at the 
quantum gravity level. Still there are questions that can be pursued 
without any quantum theory at hand. Thus we can observe that there are 
many thermodynamical systems, but---due to uniqueness theorems---only 
a few kinds of stationary black holes. Hence the Gibbs surfaces that 
arise in black hole thermodynamics are of necessity very special. We 
can ask: in what way? 

In this paper we will employ a device that was introduced in 
thermodynamics by Weinhold \cite{Weinhold} and Ruppeiner 
\cite{Ruppeiner}. Their observation is that 
the Hessian matrix of the second derivatives of the energy, or alternatively 
the entropy, can be regarded as a Riemannian metric on the space of 
thermodynamical states. When energy is used as a potential this metric is 
called the Weinhold metric, when the entropy is used it is called the 
Ruppeiner metric. It is essential that the energy and the entropy are 
regarded as functions of the extensive variables, such as volume and 
particle number. Ruppeiner's proposal is related to the use of the 
canonical ensemble, and his metric is closely connected to 
thermodynamic fluctuation theory. For self--gravitating systems it is 
natural to work with the microcanonical ensemble, and extensivity does 
not hold. But we can still demand that energy and entropy are functions 
of mechanically conserved control parameters, such as angular momentum 
and electric charge, and then proceed as before. The interesting thing 
is that very special metrics ensue---the Weinhold metric turns out to be flat 
for the Kerr black hole in arbitrary spacetime dimension (provided that 
only one spin parameter is used), while the Ruppeiner metric is flat 
for the Reissner--Nordstr\"om black hole in arbitrary spacetime dimension 
\cite{ABP} \cite{AP}. This is so provided that the cosmological constant 
is kept to zero. When a (negative) cosmological constant is turned on 
the thermodynamical metrics develop curvature \cite{ABP}. (The 2+1 dimensional 
BTZ black hole is an exception; it has a flat Ruppeiner metric.)  

There is more to this story. There are numerous studies 
of soluble models in statistical mechanics which suggest that the 
detailed behaviour of the curvature of the Ruppeiner metric carries 
information about phase transitions, and indeed about the underlying 
statistical mechanical model (see Ruppeiner \cite{Ruppeiner}, and a more 
recent review 
\cite{Des}). Something similar may be true in black hole thermodynamics; 
in particular Arcioni and Lozano-Tellechea argue that it is relevant 
for fluctuations around near extremal black holes \cite{Arcioni}. 
Let us also take note of the suggestion that the thermodynamical metrics 
may provide us with aspects of black hole physics that are safe against 
quantum corrections \cite{Ferrara}. In a different direction we observe 
that the use of a Hessian matrix as a Riemannian metric arises in other 
contexts. The obvious example is mathematical statistics, 
where such metrics are known as information metrics. What thermodynamics 
and mathematical statistics have in common---the connections at the level 
of statistical mechanics apart---is a preferred affine structure with respect 
to which the second derivatives are defined. 

The present paper has two purposes. First, to investigate the mathematical 
requirements for having a flat Ruppeiner metric. The conclusion is that the 
black hole examples have flat thermodynamic geometries, with wedge shaped 
state spaces, because of the special quasi--homogeneity properties of their 
fundamental relations. A second purpose is to explore the picture of state 
space offered by the Ruppeiner theory, and incidentally to comment on some 
criticism directed against our earlier work \cite{Cai}.  

\vspace{5mm}

{\bf 2. Flat information metrics}

\

\noindent We begin with some generalities. We study metrics that are defined, 
in some preferred affine coordinate system, by 

\begin{equation} g_{ij} = \partial_i\partial_j\psi \ . \label{1} \end{equation} 

\noindent The potential $\psi$ can be any reasonable function. Examples 
occur in mathematical statistics, where a favoured choice of 
potential is 

\begin{equation} \psi = \sum_{i = 1}^Nx^i \ln{x^i} \ , \hspace{5mm} x^i > 0 \ . 
\end{equation}

\noindent As it stands this is a flat metric on the positive cone, but if 
the condition that the positive numbers $x^i$ sum to unity is imposed it 
becomes round---and the potential becomes equal to the Shannon entropy with 
sign reversed, while the metric itself is known as the Fisher information 
matrix  \cite{Amari}. This serves to explain why such metrics are called 
information metrics.   

In thermodynamics the potential is either the entropy with sign 
reversed, or the energy function. If 

\begin{equation} \psi = - S(M,Q) \end{equation}

\noindent the corresponding metric is known as the Ruppeiner metric, if 

\begin{equation} \psi = M(S,Q) \end{equation} 

\noindent it is the Weinhold metric. These two metrics are related by a 
conformal factor equal to the temperature, 

\begin{equation} ds^2_W = Tds^2_R \ , \hspace{8mm} T \equiv \left( 
\frac{\partial M}{\partial S}\right)_Q \ . \label{conform} \end{equation}

\noindent The preferred affine coordinate systems 
are provided by the mechanically conserved control parameters, including 
energy or entropy. Our choice of the letter $M$ for the energy is of course 
dictated by our interest in black hole thermodynamics. For the 
Reissner--Nordstr\"om black hole $Q$ denotes the electric charge, while for 
Kerr it stands in for angular 
momentum. In this paper we will consider two dimensional state spaces only. 

Our question is: when is an information metric flat? One possibility is 
that 

\begin{equation} \psi = \sum_{i=1}^Nf_i(x^i) \ . \end{equation}

\noindent The Fisher metric on the positive cone comes from such a potential. 
Moreover the Ruppeiner metric of the ideal gas at fixed particle number 
is of this type. However, the black hole information metrics are not. 

Next consider potentials that have the quasi--homogeneity property 

\begin{equation} \lambda^{a_3}\psi(x,y) = \psi (\lambda^{a_1}x, 
\lambda^{a_2}y) \ . \end{equation}

\noindent We can afford to assume that $x > 0$ and $\psi > 0$. The 
property is equivalent to 

\begin{equation} \psi(x,y) = x^af(x^by) \ , \label{pot} \end{equation}

\noindent where $f$ is some function and $a$, $b$ are some exponents. (To 
show this \cite{Hankey}, choose $\lambda^{a_1}x = 1$. One finds 
$a = a_3/a_1$, $b = - a_2/a_1$.) Note that 

\begin{equation} S(M,Q) = M^af(M^bQ) \hspace{5mm} \Leftrightarrow 
\hspace{5mm} M = S^{1/a}h(S^{b/a}Q) \ . \end{equation}

\noindent We can now state a small theorem, namely: 
If $\psi(x,y) = x^af(y/x)$ then the information metric is flat. 
The converse does not hold. 

The most straightforward proof of this theorem is to change to the new 
coordinates

\begin{equation} \psi = x^af(x^by) \hspace{4mm} \mbox{and} \hspace{4mm} 
\sigma = x^by \ . \end{equation}

\noindent (To avoid misunderstanding: the metric is always defined using 
differentiation with respect to the preferred coordinates---but once the 
metric is given we can use any coordinates 
we please.) An explicit calculation shows that

\begin{eqnarray} ds^2 = \left( \frac{a-1}{a} - \frac{b(b+1)}{a^2} 
\frac{\sigma f^\prime}{f}\right) \frac{d\psi^2}{\psi} + 
2(b+1)\left( \frac{1}{a}\frac{f^\prime}{f} + \frac{b}{a^2}
\frac{\sigma f^{\prime 2}}{f^2}\right) d\psi d\sigma + \nonumber \\
\ \\
\hspace{1cm} + \psi 
\left( \frac{f^{\prime \prime}}{f} - \frac{2b + a + 1}{a}\frac{f^{\prime 2}}
{f^2} -  \frac{b(b+1)}{a^2}\frac{\sigma f^{\prime 3}}
{f^3}\right) d\sigma^2 \ . \nonumber \label{11} \end{eqnarray}

\noindent This is diagonal if $b = -1$. If we introduce the new coordinate 
$r = \sqrt{\psi}$ it is also manifestly a flat metric, and it covers a 
wedge shaped region. Given the function $f$ we can reparametrize $\sigma$ so 
that we end up with polar coordinates, or Rindler coordinates if the metric is 
Lorentzian, and read off the opening angle of the wedge. Anyway the small 
theorem is proved. There is an exception if $b = - 1$ and $a = 1$, since 
then the metric is 
degenerate. This is so for homogeneous potentials in any dimension. 

An intermediate step in the calculation is of interest as well. Using 
$x$ and $\sigma$ as coordinates we get

\begin{equation} ds^2 = x^{a-2}\Bigl( (a(a-1)f - b(b+1)\sigma f^\prime )dx^2 
+ 2(a+b)xf^\prime dxd\sigma + x^2f^{\prime \prime}d\sigma^2\Bigr) \ . \end{equation}

\noindent If $a + b = 0$ this is diagonal, and can be written as 

\begin{equation} ds^2 = \psi_{,x}\left( (a-1)\frac{dx^2}{x} + \frac{x}{a}
\frac{f^{\prime \prime}}{f-\sigma f^{\prime}}d\sigma^2 \right) \ , \end{equation}

\noindent where $\psi_{,x}$ is the derivative with respect to $x$ of $\psi(x,y)$. 
This is the metric on a flat wedge multiplied with the conformal factor 
$\psi_{,x}$, and should be compared to eq. (\ref{conform}).  

We can restate the small theorem in thermodynamical language: Let 

\begin{equation} S = M^af(M^bQ) \ . \end{equation}

\noindent If $b = - 1$ the Ruppeiner metric is flat. If $a+b = 0$ the 
Weinhold metric is flat.

It is instructive to look at the Riemann curvature tensor as well. 
In the preferred coordinate system the Christoffel symbols (with one index 
lowered using the metric) are given by 

\begin{equation} \Gamma_{ijk} = \frac{1}{2}\partial_i\partial_j
\partial_k\psi \ . \label{skew} \end{equation} 

\noindent In mathematical statistics this is also known as the skewness 
tensor \cite{Amari}. The expression for the Riemann curvature tensor 
then simplifies to 

\begin{equation} R_{ijkl} = \Gamma_{ikm}g^{mn}\Gamma_{njl} 
- \Gamma_{ilm}g^{mn}\Gamma_{njk} \ . \label{Frob} \end{equation} 

\noindent For our special choice of potential, eq. (\ref{pot}), setting 
the Riemann tensor to zero results in a non--linear third order ODE of 
somewhat frightening aspect. To be precise about it, the curvature scalar is 

\begin{eqnarray} R = \frac{(b+1)x^{3a + 4b - 4}}{2g^2} 
[a(a-1)(a+b)ff^{\prime}f^{\prime \prime \prime}- 
2a(a-1)(a+2b)ff^{\prime \prime 2} - \nonumber \\
\ \nonumber \\
 -ab(a-1)\sigma ff^{\prime \prime}f^{\prime \prime \prime} + 
(a+b)^2(a+b-1)f^{\prime 2}f^{\prime \prime} + \nonumber \\
\ \\ 
+ b(a+b)(2a+b-1)\sigma f^{\prime 2}f^{\prime \prime \prime} + \nonumber \\
\ \nonumber \\
+b(2b -a^2-3ab)\sigma f^{\prime}f^{\prime \prime 2} + 
b^2(b+1)\sigma^2(f^{\prime}f^{\prime \prime}f^{\prime \prime \prime} -
f^{\prime \prime 3})] \nonumber \end{eqnarray} 

\noindent where $g$, the determinant of the metric, is 

\begin{equation} g = x^{2(a + b -1)}[a(a-1)ff^{\prime \prime} - 
(a+b)^2f^{\prime 2} - b(b+1)\sigma f^{\prime}f^{\prime \prime}] \ . 
\end{equation} 

\noindent For our present purposes the main feature of $R$ is that it 
has $(b+1)$ as a prefactor, and therefore the metric is flat for $b = - 1$ 
whatever the form of the function $f$. 

Regardless of why it happens, flatness has an interesting mathematical 
consequence. Suppose that the Riemann tensor vanishes. A somewhat similar 
structure arises in the theory of Frobenius manifolds \cite{Dubrovin}, 
where it is observed that the resulting equation can be used to define 
an algebra through

\begin{equation} \partial_i \circ \partial_j = {\Gamma}_{ijm}g^{mk}\partial_k 
\ . \end{equation}

\noindent This algebra is commutative by construction, and associative because 
of eq. (\ref{Frob}). Such algebras are used to describe the moduli space 
of topological conformal field theories; this sounds as if it might, through 
some back door, have some connection to black hole thermodynamics, but in 
fact the two settings are very different. In the theory of Frobenius manifolds 
the metric tensor used in eq. (\ref{Frob}) is a fixed quadratic form, 
and the skewness tensor (\ref{skew}) is not a Christoffel symbol of any 
relevant metric. Thus, reluctantly, we conclude that the theory of Frobenius 
manifolds is irrelevant to us.

\vspace{5mm} 

{\bf 3. Black hole examples}

\

\noindent For black holes the fundamental relation relates the area of 
the event horizon to the ADM charges of the black hole. More precisely 
we set $S = kA/4$, where $A$ is the area of the event horizon and $k$ 
is Boltzmann's constant. We adjust the numerical value of the latter to 
simplify the resulting expression. 

When the cosmological constant vanishes, the Einstein-Maxwell equations are 
scale invariant. This has consequences for the solutions, which can be 
deduced by dimensional analysis. Using length as the only basic unit, the 
black hole control parameters have dimensions
 
\begin{equation} [S] = L^{d-2} \ , \ \ [M] = L^{d - 3} \ , \ \ 
[Q] = L^{d-3} \ , \ \ [J] = L^{d-2} \ , \end{equation}

\noindent where $d$ is the dimension of spacetime. It follows that the 
fundamental relation will have quasi-homogeneity 
properties, with definite exponents. Indeed 

\begin{equation} L^{d-2}S(M,Q,J) = S(L^{d-3}M, L^{d-3}Q, L^{d-2}J) \ . 
\end{equation}

\noindent Hence, by the result quoted in the previous section, in the 
two parameter cases the fundamental relations must be 

\begin{equation} S = M^{\frac{d-2}{d-3}}f\left( \frac{Q}{M}\right) \hspace{5mm} 
\mbox{and} \hspace{5mm} S = 
M^{\frac{d-2}{d-3}}f\left( \frac{J}{M^{\frac{d-2}{d-3}}}\right) \ . \end{equation}

\noindent Finally the theorem proved in section 2 implies that the Ruppeiner 
geometry of the Reissner--Nordstr\"om black holes will be flat in any dimension, 
and similarly the Weinhold geometry of the Kerr black holes is flat 
in any dimension. This will be true also for ``exotic'' Kerr black holes 
such as the ``black ring'' in five dimensions \cite{Emparan}. 

Some explicit examples are as follows. The Reissner--Nordstr\"om black hole 
in arbitrary spacetime dimension $d$ has the fundamental relation

\begin{equation} S = M^c\left( 1 + \sqrt{1 - \frac{c}{2}\frac{Q^2}{M^2}}
\right)^c \ , \hspace{6mm} c \equiv \frac{d-2}{d-3} \ . \end{equation}

\noindent The Ruppeiner geometry is  
a timelike wedge in a flat Minkowski space, with an opening angle that 
grows with $d$. It is a black hole if the integer $d \geq 4$. The Kerr 
black hole in spacetime dimension $d$ has the fundamental relation 

\begin{equation} M = \frac{d-2}{4}S^{\frac{d-3}{d-2}}\left( 1 + 
\frac{4J^2}{S^2}\right)^{1/(d-2)} \ . \label{Kerr} \end{equation}

\noindent The Weinhold geometry is a timelike wedge in Minkowski space 
for $d = 4$ and $d = 5$, while it fills the entire forwards light cone 
when $d \geq 6$ (due to the absence of extremal Kerr--Myers--Perry black 
holes in these dimensions \cite{Myers}). An explicit form of the fundamental 
relation for the black ring can be found in the literature \cite{Arcioni}. 
For the three dimensional Kerr--Newman family 
the Ruppeiner and Weinhold metrics are both curved \cite{ABP}.

Our dimensional argument fails in the presence of a cosmological constant. 
In spite of this the 2+1 dimensional BTZ black hole \cite{Banados} has a 
fundamental relation of the form $S = M^af(J/M)$, and hence its Ruppeiner 
state space is a flat wedge (in an Euclidean space) \cite{ABP}. In higher 
dimensions anti-de Sitter black holes have curved thermodynamic geometries.  

It is instructive to compare the black hole examples to the ideal gas, 
which has the fundamental relation

\begin{equation} S = N\ln{\left[ \frac{V}{N}\left( \frac{U}{N}\right)^c 
\right]} + k_1N \hspace{5mm} \Leftrightarrow \hspace{5mm} U = k_2
\frac{N^{(c+1)/c}}{V^{1/c}}e^{S/(cN)} 
\ , \end{equation}

\noindent where $c$ is the ratio of specific heats, $k_1$ and $k_2$ 
are constants, and we use $U$ for energy. Since $S = S(U,V,N)$ is a 
homogeneous function the three dimensional Ruppeiner metric is actually 
degenerate, but if we consider the ideal gas at fixed volume $V$ it 
belongs to the class (\ref{pot}). For the Ruppeiner case we have 
$a + b = 0$, while for the Weinhold case $b = - 1$. Hence the Weinhold 
geometry is flat; the opening angle of its wedge turns out to go between 
zero and infinity, so it is actually an infinite covering of the 
punctured plane. But more is true: because of the quite special function 
involved the Ruppeiner metric also is flat. Similarly both the Weinhold and 
Ruppeiner geometries of the ideal gas at fixed particle number $N$ are 
flat; the latter describes a flat plane \cite{Nulton}. This illustrates 
that our small theorem gives a sufficient but not necessary condition 
for flatness, and it shows that the ideal gas is even more special than 
our black hole examples. Another case where both the Weinhold and the 
Ruppeiner metrics are flat is given by the Kerr formula (\ref{Kerr}), 
for the unphysical value $d = 3$. 
 
\vspace{5mm}
 
{\bf 4. The Reissner--Nordstr\"om black hole}

\

\noindent In this section we focus on the Reissner--Nordstr\"om family of 
black holes in four spacetime dimensions. The Gibbs surface is defined by 
the fundamental relation

\begin{equation} S = M^2\left( 1 + \sqrt{1 - \frac{Q^2}{M^2}}\right)^2 \ 
. \end{equation}

\noindent The Ruppeiner metric can be obtained from eq. (\ref{11}). Actually 
it is convenient to trade the coordinate $\sigma = Q/M$ for 

\begin{equation} u \equiv \frac{\sigma}{1 + \sqrt{1-\sigma^2}} = 
\frac{Q}{\sqrt{S}} = \left( \frac{\partial M}{\partial Q}\right)_S \ , 
\hspace{6mm} - 1 \leq u \leq 1 \ . \end{equation} 

\noindent This coordinate is conjugate to the charge, and equals the 
electric potential at the event horizon. We now get the metric in the form 

\begin{equation} ds^2 = - \frac{dS^2}{2S} + \frac{4Sdu^2}{1-u^2} = 
- d\tau^2 + \tau^2d\chi^2 \ . \end{equation}

\noindent In the last step we traded our coordinates for 

\begin{equation} \tau = \sqrt{2S} \hspace{10mm} \chi = \sqrt{2}\arcsin{u} \ . 
\end{equation}

\noindent The coordinates $\tau$ and $\chi$ are the usual Rindler coordinates 
on the forward lightcone in Minkowski space. If we like we can introduce the 
inertial coordinates 

\begin{equation} t = \tau \cosh{\chi} \hspace{9mm} x = \tau \sinh{\chi} \ . 
\end{equation}

\noindent A picture of the state space 
as a flat wedge is given in Fig. 1, with some details added. Our 
picture for the Kerr case (based on the Weinhold metric) is qualitatively 
similar, although the wedge is thinner and curves of constant $Q$ are 
replaced by curves of constant $J$. 

\begin{figure}
\centering

\begin{tabular}{cc}

\epsfig{file=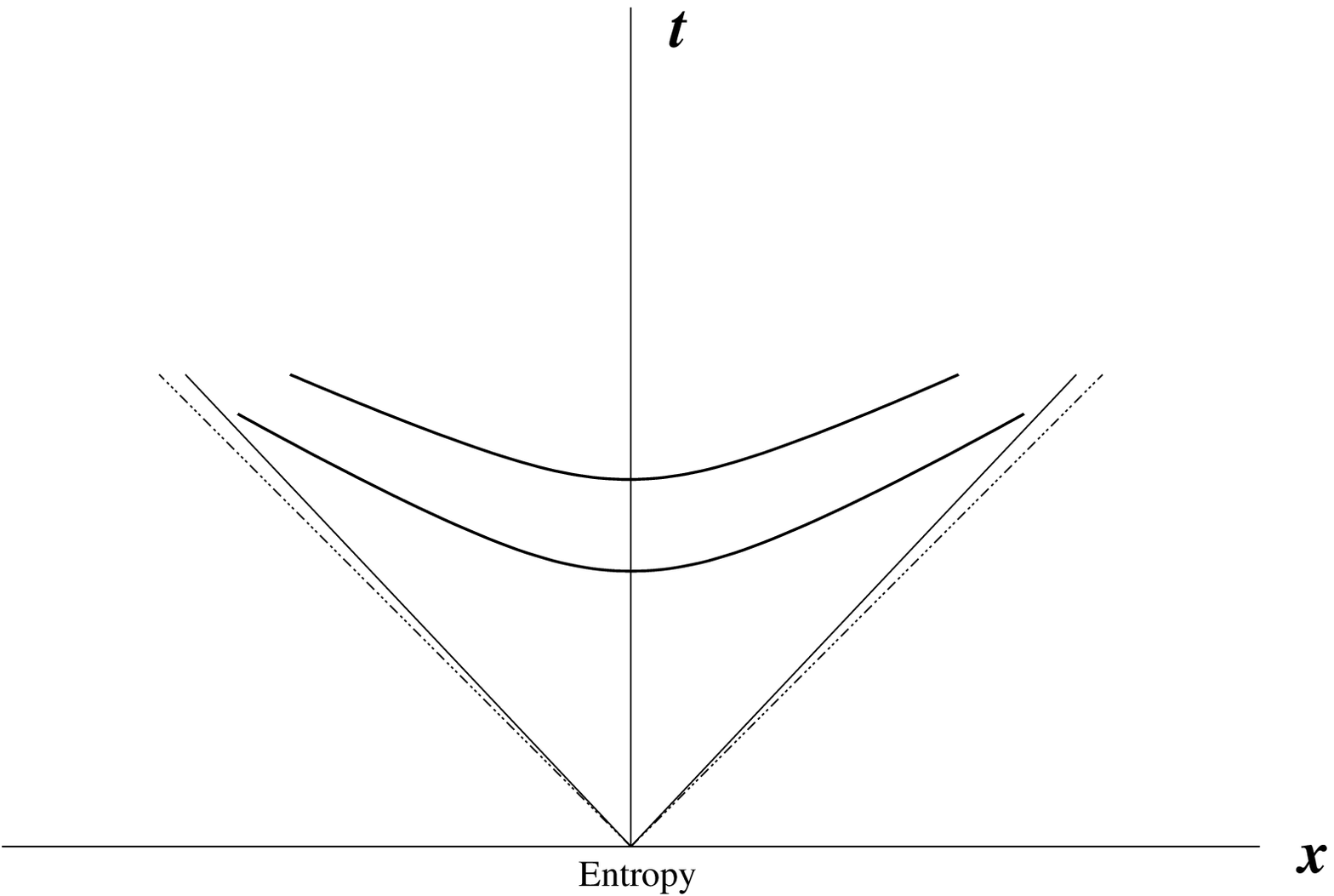,width=0.46\linewidth} & 
\epsfig{file=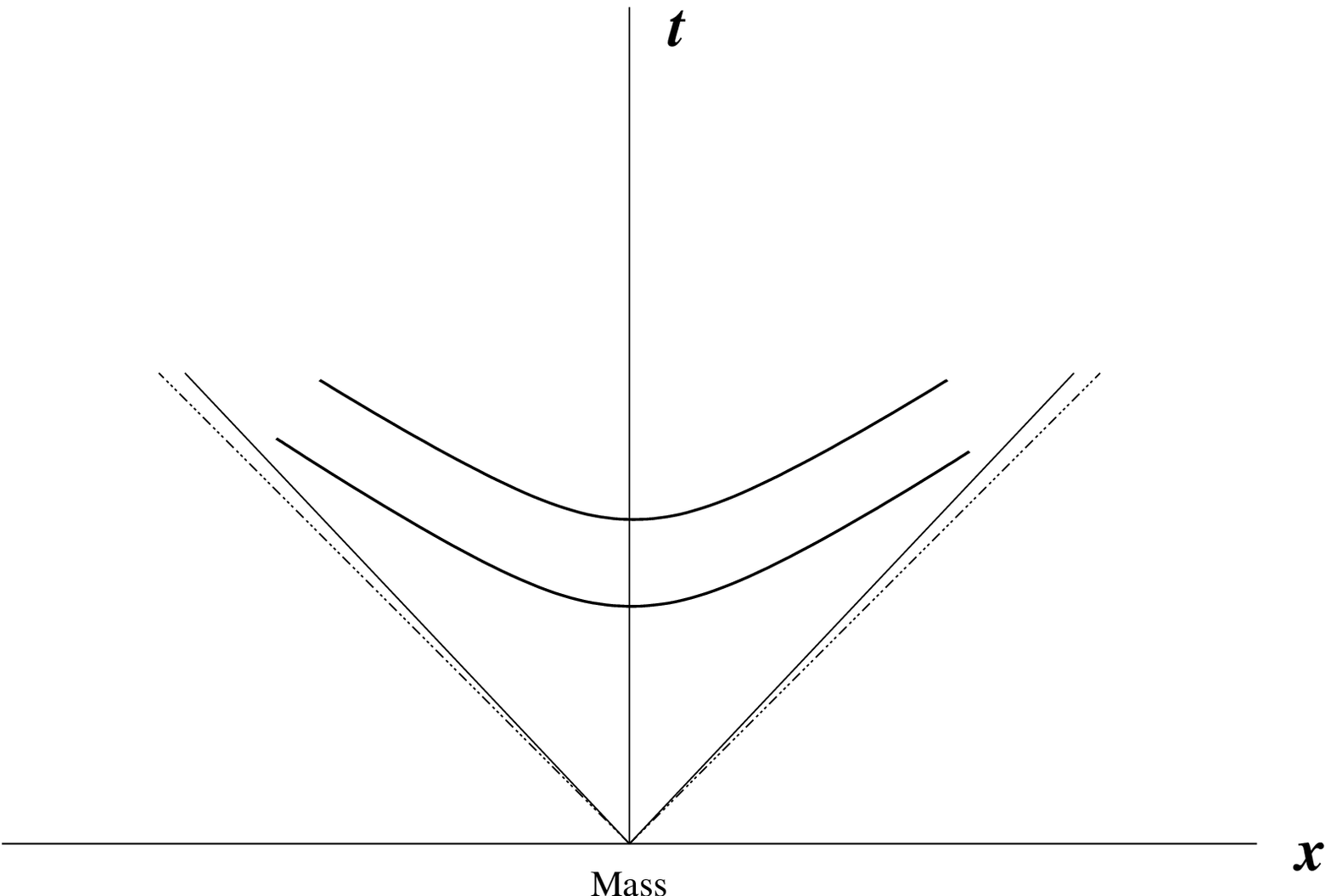,width=0.46\linewidth} \\[0.5cm]
\epsfig{file=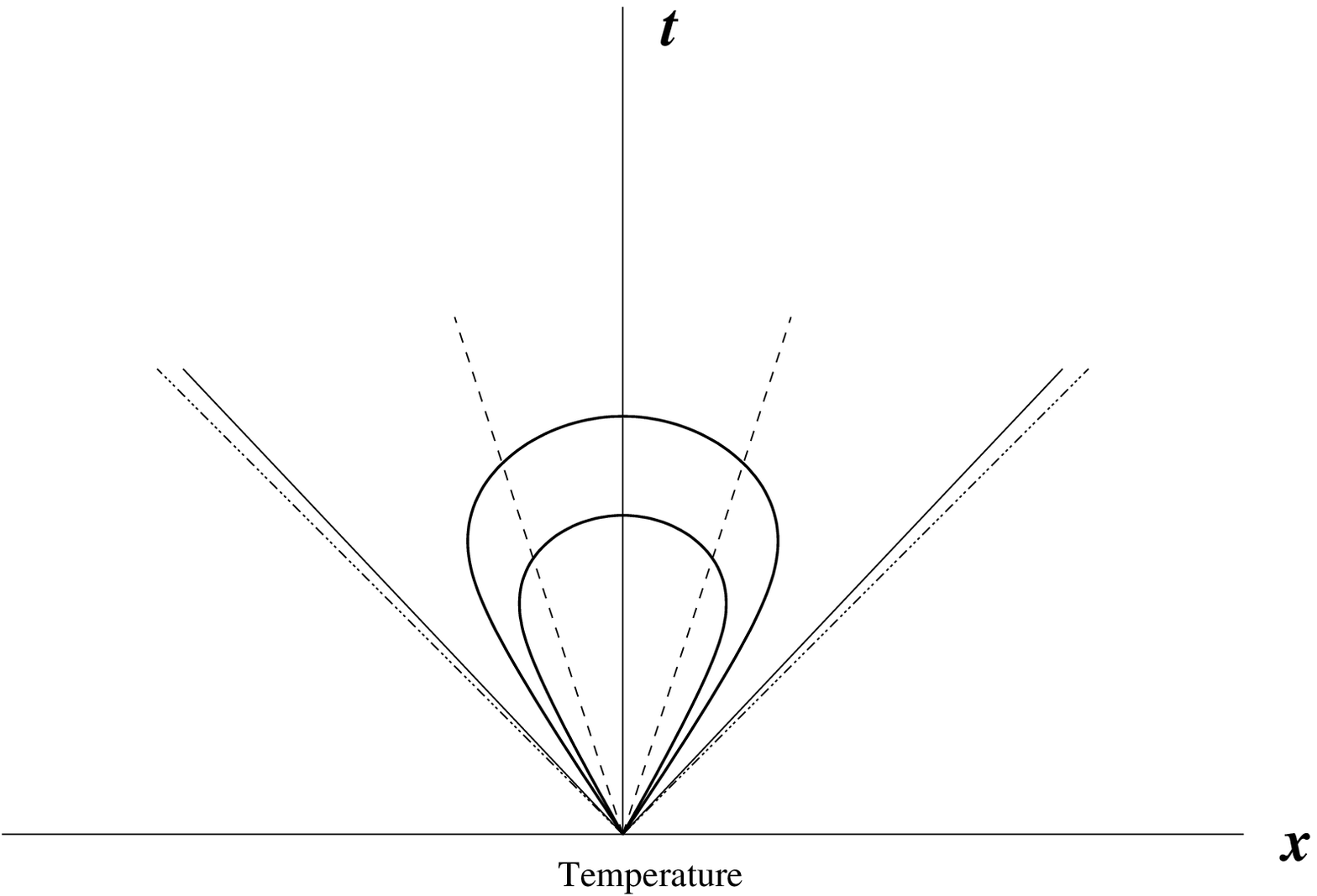,width=0.46\linewidth} & 
\epsfig{file=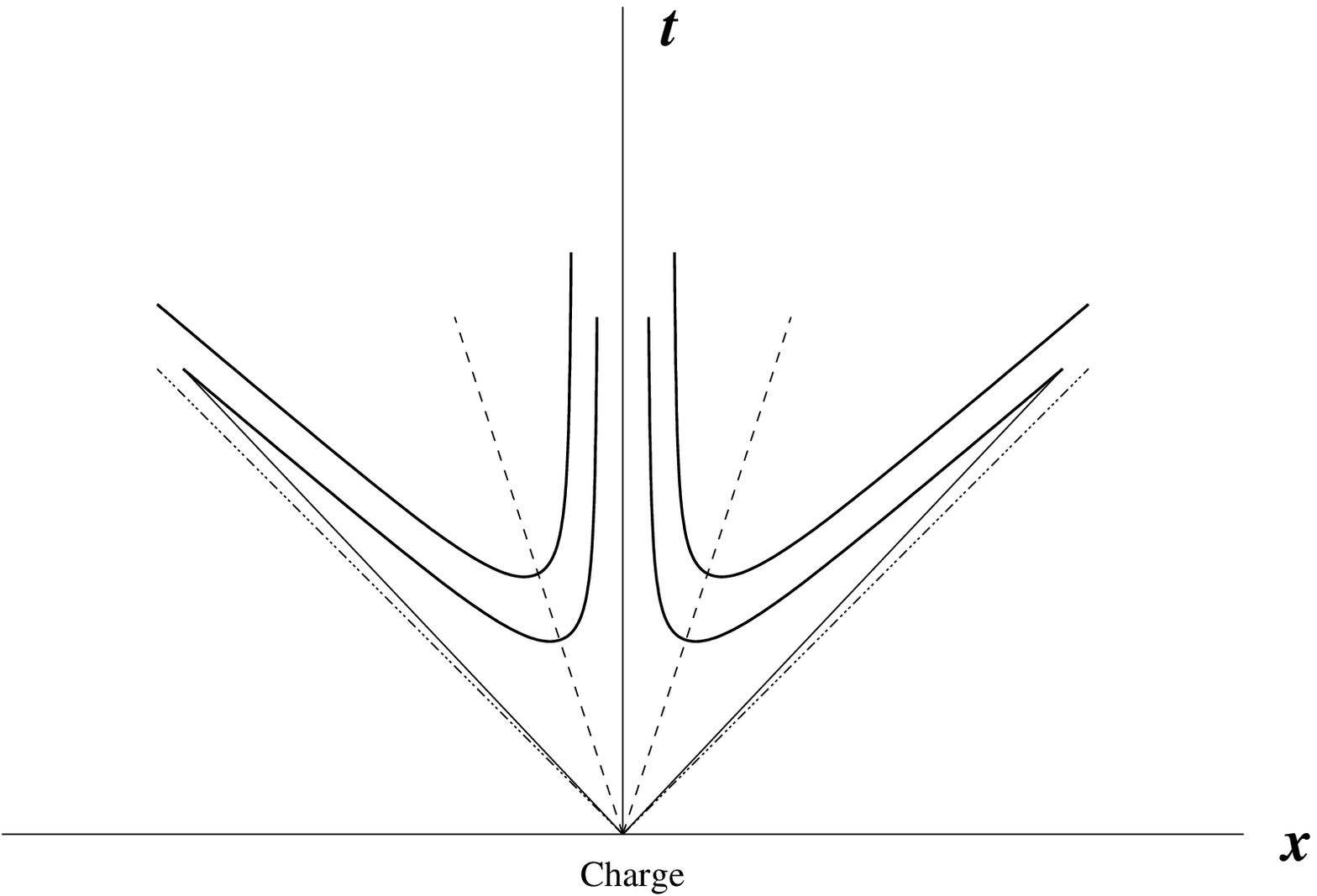,width=0.46\linewidth} 

\end{tabular}
\caption{\small The state space of the Reissner-Nordstr\"om black hole 
is a wedge 
inside the forwards light cone of a 1+1 dimensional Minkowski space. We show 
curves of constant entropy (spacelike hyperbolas), constant mass (also 
spacelike), constant temperature, and constant charge. The latter two become 
null at the ``Davies point'', which is given by a dashed line of constant 
electric potential.}

\end{figure}

In a recent paper it was argued that the entropy ought to be expressed 
as a function of the enthalphy and the electric potential, before the 
derivatives are taken \cite{Cai}. We do not wish to appear dogmatic on 
this, or any other point. But we are concerned with the consequences of the 
definitions that we have stated. There was also a more specific 
criticism. It was argued (long ago) by Davies \cite{Davies} that charged 
black holes suffer a second order phase transition at $Q/M = \sqrt{3}/2$. 
The argument was based on the observation that the specific heat

\begin{equation} c_Q \equiv T\left( \frac{\partial S}{\partial T}\right)_Q 
= \frac{2S(S-Q^2)}{3Q^2 - S} \end{equation}

\noindent diverges there, and then changes sign. Although this argument 
was quickly challenged \cite{Tranah}, it resurfaces now and then. Thus 
Penrose \cite{Penrose} has used it to suggest that it should 
be qualitatively easier to understand the entropy of an extremal black 
hole in state counting terms, because unlike its Schwarzschild counterpart 
such a black hole has positive specific heat. It has also been used to suggest 
that the Ruppeiner metric as defined by us must be irrelevant since ``a 
statistical model without any interaction cannot reproduce thermodynamic 
properties of the RN black hole'' \cite{Cai}. Now we never claimed that the flat 
Ruppeiner metric proves that the ``statistical model'' is non--interacting. 
Nor are we worried by any phase transition at $Q/M = \sqrt{3}/2$. As explained 
by Sorkin \cite{Sorkin} and others \cite{Katz}, a microcanonical 
instability would occur only if the specific heat changed sign through zero.  

From our present point of view let us observe that the specific heat is not 
only a property of the Gibbs surface, it is also a property of the special 
curve along which we evaluate the specific heat. As one can see in the 
picture, what happens at $Q/M = \sqrt{3}/2$ is that the particular curve 
defined by constant $Q$ changes from timelike (negative $c_Q$) 
to spacelike (positive $c_Q$). At $Q/M = \sqrt{3}/2$ the curve is null, 
which means that the heat capacity related to this curve diverges 
\cite{Weinhold}. Similarly, curves of constant $J$ become null at the 
``Davies' point'' of the Kerr black hole. Indeed this kind of behaviour 
will occur, for some curves, in every point on a Gibbs surface where the entropy 
function is non-concave---as it typically will be for self-gravitating 
systems \cite{Lynden}. For the question of stability in the microcanonical 
ensemble this is quite irrelevant.

\vspace{5mm}

{\bf 5. Conclusions}

\

\noindent We have investigated the fact that some black hole families 
have either flat Ruppeiner metrics or flat Weinhold metrics. We also 
used this fact to draw a simple picture of the state space of 
Reissner--Nordstr\"om black holes. 

Our main result is that eqs. (\ref{1}) and (\ref{pot}), 
with $b = -1$, always give a flat information metric defined on a wedge 
in a flat space. All our black hole examples belong to this class; in 
the asymptotically flat case this is a consequence of the scale invariance 
of the Einstein equations. It is perhaps a little disappointing that we 
do not obtain any restriction on the free function contained in the 
potential. 

Although we have understood why certain thermodynamical metrics of some 
black holes are flat, there are many questions that remain to be investigated. 
There may be more to the fact that Reissner-Nordstr\"om black holes have 
a flat Ruppeiner geometry, while that of the Kerr black holes is curved. 
After all for the former there are quantum gravity based calculations of 
the entropy, but not so far for the latter. The broader question 
about the nature of the Gibbs surfaces that appear in black hole thermodynamics 
has many aspects; in this paper we have investigated only one of them.  

\vspace{1cm}

{\bf Acknowledgements}

\

\noindent IB thanks Sergei Merkulov for his explanation of Frobenius 
manifolds, and the VR for support. NP is grateful to his parents for 
financial support, and would like to thank Joakim Sj\"ostrand and 
John Ward for useful discussions on plots.

We dedicate this paper to Rafael Sorkin's 60th birthday.

\end{document}